# Multiplication of freestanding semiconductor membranes from a single wafer by advanced remote epitaxy

Hyunseok Kim[1,2,†], Yunpeng Liu[1,2,†], Kuangye Lu[1,2,†], Celesta S. Chang[1,2,†], Kuan Qiao[1,2], Ki Seok Kim[1], Bo-In Park[1,2], Junseok Jeong[1,2], Menglin Zhu[3], Jun Min Suh[1,2], Yongmin Baek[4], You Jin Ji[5], Sungsu Kang[6,7], Sangho Lee[1,2], Ne Myo Han[1,2], Chansoo Kim[1,2], Chanyeol Choi[1,2,8], Xinyuan Zhang[1,9], Haozhe Wang[8], Lingping Kong[9], Jungwon Park[6,7], Kyusang Lee[4], Geun Young Yeom[5,10], Sungkyu Kim[11], Jinwoo Hwang[3], Jing Kong[8], Sang-Hoon Bae[12,13], Wei Kong[14], Jeehwan Kim[1,2,9,*]

[1]Research Laboratory of Electronics, Massachusetts Institute of Technology, Cambridge, MA 02139, USA
[2]Department of Mechanical Engineering, Massachusetts Institute of Technology, Cambridge, MA 02139, USA
[3]Department of Materials Science and Engineering, The Ohio State University, Columbus, OH 43210, USA
[4]Department of Electrical and Computer Engineering, University of Virginia, Charlottesville, VA 22904, USA
[5]School of Advanced Materials Science and Engineering, Sungkyunkwan University, Suwon 16419, Republic of Korea
[6]School of Chemical and Biological Engineering, Seoul National University; Seoul 08826, Republic of Korea
[7]Center for Nanoparticle Research, Institute for Basic Science (IBS), Seoul 08826, Republic of Korea
[8]Department of Electrical Engineering and Computer Science, Massachusetts Institute of Technology, Cambridge, MA 02139, USA
[9]Department of Materials Science and Engineering, Massachusetts Institute of Technology, Cambridge, MA 02139, USA
[10]SKKU Advanced Institute of Nano Technology (SAINT), Sungkyunkwan University, Suwon 16419, Republic of Korea
[11]HMC, Department of Nanotechnology and Advanced Materials Engineering, Sejong University, Seoul 05006, Republic of Korea
[12]Department of Mechanical Engineering and Materials Science, Washington University in Saint Louis, St. Louis, MO 63130, USA
[13]Institute of Materials Science and Engineering, Washington University in Saint Louis, St. Louis, MO 63130, USA
[14]Department of Materials Science and Engineering, Westlake University, Hangzhou 310024 Zhejiang, China

[†]These authors contributed equally.
[*]Correspondence to jeehwan@mit.edu

**Abstract**

Freestanding single-crystalline membranes are an important building block for functional electronics. Especially, compounds semiconductor membranes such as III-N and III-V offer great opportunities for optoelectronics, high-power electronics, and high-speed computing. Despite huge efforts to produce such membranes by detaching epitaxial layers from donor wafers, however, it is still challenging to harvest epitaxial layers using practical processes. Here, we demonstrate a method to grow and harvest multiple epitaxial membranes with extremely high throughput at the wafer scale. For this, 2D materials are directly formed on III-N and III-V substrates in epitaxy systems, which enables an advanced remote epitaxy scheme comprised of multiple alternating layers of 2D materials and epitaxial layers that can be formed by a single epitaxy run. Each epilayer in the multi-stack structure is then harvested by layer-by-layer peeling, producing multiple freestanding membranes with unprecedented throughput from a single wafer. Because 2D materials allow peeling at the interface without damaging the epilayer or the substrate, wafers can be reused for subsequent membrane production. Therefore, this work represents a meaningful step toward high-throughput and low-cost production of single-crystal membranes that can be heterointegrated.

In recent years, the need for freestanding single-crystalline membranes of functional materials is gaining huge attention because the freestanding form can provide extra degrees of freedom in their implementations. For example, heterogenous integration of dissimilar materials, which is not feasible by conventional heteroepitaxy due to the incompatibility between materials' crystal structures and the harshness of the epitaxy environment, can be realized by stacking freestanding membranes[1]. By decoupling various types of epitaxial layers from donor wafers and integrating them, the physical properties of dissimilar materials can be easily coupled, which are unclamped from the substrate and can be made flexible and lightweight[2]. More importantly, the adoption of freestanding membranes instead of bulky wafers provides significant cost savings for the production of devices. Especially, for non-silicon materials, the cost of wafers is often orders of magnitude higher than that of silicon, limiting the utilization of non-silicon materials for device applications. So far, a variety of lift-off techniques including chemical, optical, and mechanical lift-off, have been developed for producing single-crystalline freestanding membranes and for reusing the costly wafer after the lift-off. However, all of these lift-off methods damage or contaminate the substrate during the process, necessitating a thorough refurbishing step such as chemical-mechanical polishing (CMP). This not only substantially adds the cost, but also limits the number of lift-off processes that could be conducted from a single wafer by recycling the wafer. Furthermore, conventional lift-off methods are applicable only to specific material systems. The chemical lift-off process, for example, requires a lattice-matched sacrificial layer with etching selectivity, such as AlAs sacrificial layer in GaAs active layer[3]. Moreover, chemical lift-off is typically an extremely slow process taking more than several hours, leaves behind byproducts that require refurbishing processes, and prevents the use of the sacrificial material in the device layer[3,4]. Optical lift-off requires a transparent substrate to selectively melt the interface by lasers, wherein the epilayer and the substrate near the interface are severely damaged by heat[5], and the mechanical lift-off process significantly roughens the substrate by spalling[6,7]. Thus, it is crucial to develop a process that can produce multiple single-crystalline membranes without such limitations to benefit from wafer recycling.

2D material-based layer transfer (2DLT) process, which employs 2D van der Waals (vdW) materials as an interfacial layer, allows ultrafast mechanical lift-off precisely at the 2D interface, due to a very weak adhesion between the vdW layer with the overlayer[8,9]. The overlayer can be epitaxially grown by remote epitaxy using graphene as a 2D interlayer, which provides a new path toward single-crystalline membrane production[9–12]. In current technology, however, graphene is typically transferred from foreign templates onto the substrates of interest, which inevitably induces defects in graphene during the transfer process, such as tears, wrinkles, and residues. This not only deteriorates the crystal quality of membranes grown by remote epitaxy, but incurs local spalling from defective areas of graphene during 2DLT, damaging both the substrate and the exfoliated epilayer[9]. Furthermore, because the host wafer does not remain atomically smooth after 2DLT due to such damages, refurbishing processes are still necessary to attain an epi-ready surface to reuse the wafer, like other lift-off techniques.

Here, we demonstrate a practical and universal method to grow and harvest epitaxial membranes with extremely high throughput at the wafer scale. We develop an advanced remote epitaxy technique to form multiple alternating layers of 2D materials and epitaxial films by a single growth run, where the epilayers are mechanically exfoliated by a layer-by-layer lift-off process, producing multiple freestanding membranes. After epilayers are harvested by 2DLT, the wafer can be reused to repeat the process and keep producing freestanding membranes. Therefore, this approach will truly and substantially reduce the cost of non-Si electronics by high-throughput production of semi-infinite copies of single-crystalline membrane from a single wafer. Such an unprecedented method is realized by the following key developments. First, we demonstrate a method to grow 2D materials directly on compound semiconductor wafers, such as III-N and III-V, in epitaxy equipment at low temperatures so that surface dissociation can be avoided. In a III-V metal-organic chemical vapor deposition (MOCVD) reactor, toluene is employed as a carbon precursor with a low cracking temperature to form graphene on III-V, and in a molecular-beam epitaxy (MBE) reactor, elemental boron and nitrogen plasma are employed as precursors to form hexagonal boron nitride (h-BN) on III-N. Because of the low growth temperatures, the crystal phases of graphene and h-BN are not single-crystalline but amorphous or nanocrystalline, which still effectively support remote epitaxy. Second, multiple stacks of remote epitaxial films with 2D interlayers, such as multiple III-V/graphene stacks and multiple III-N/h-BN stacks, can be grown by a single growth run, because the growth of 2D and 3D layers share the same epitaxy tool. The III-V and III-N layers on wafers are then harvested by layer-by-layer peeling. With the Ni stressor deposited on top, exfoliation occurs from the topmost epilayer, realizing high-throughput lift-off of membranes with atomic-scale precision. Third, the wafer surface is undamaged during the peeling of epilayers, owing to the transfer-free formation of 2D layers, and thus the wafer can be reused for successive remote epitaxy and 2DLT without the need to refurbish the surface by CMP. We use the same wafer three times to repeatedly conduct remote epitaxy and 2DLT, and demonstrate three remote epitaxial layer stacks (epilayer/2D/epilayer/2D/epilayer/2D on GaAs and GaN wafers) that are consecutively detached by mechanical exfoliation at the 2D interlayers. Further multiplication of membranes from a single wafer is feasible since the quality of remote epitaxial layers and the substrates do not show any sign of degradation by stacking up the layers or by exfoliation. Thus, the proposed wafer multiplication process will open up opportunities for the electronics community to utilize materials solely by their performance without the burden of the cost of non-Si wafers. Moreover, this technique will provide a path for high-throughput production of freestanding semiconductor membranes, enabling greater flexibility in designing heterostructures as well as applications of inorganic semiconductors for conformal electronics on 3D featured surfaces.

Fig. 1a shows the process flow to harvest multiple single-crystalline semiconductor membranes from a single wafer. The process is composed of the growth of multiple sets of epilayer/2D layer, layer-by-layer exfoliation of the epilayers, and reusing the substrate to repeat the process, which maximizes the copies of membranes that can be produced from one wafer. One of the key developments to realize this high-throughput semiconductor membrane production is a transfer-free coating of 2D materials on compound semiconductor wafers by direct growth of the 2D layers in epitaxial systems which can be conducted *in situ* with remote epitaxy. One of the challenges in forming 2D vdW materials on compound semiconductor wafers is their high growth temperature, which is typically above 900 °C for graphene and h-BN. Thus, desorption from III-V and III-N surfaces occurs during the growth of 2D layers[13,14], which degrades the crystal quality of the surface and hinders the 2D layer formation. To overcome this, we develop a method to grow 2D materials on compound semiconductor wafers at a lower temperature that can minimize surface desorption. Low-temperature growth makes the crystal structure of 2D materials more defective but still with $sp^2$-dominant in-plane bonds. We first show h-BN grown directly on GaN(0001) wafers by MBE. Elemental boron in an effusion cell and nitrogen plasma are used as sources of boron and nitrogen for h-BN growth at 680 °C, which is a low enough temperature to prevent GaN desorption. After h-BN formation, the sample is taken out for characterization of low-temperature h-BN on GaN. As shown in the AFM image Fig. 1b, the GaN substrate coated with h-BN is atomically smooth with the r.m.s. roughness of 0.37 nm. X-ray photoelectron spectroscopy (XPS) measurements reveal the existence of boron and nitrogen peaks (Fig. 1c,d), although the nitrogen peak is from both the h-BN and GaN substrate with a shoulder peak at ~396 eV, which is attributed to the Auger peak of nitrogen from GaN[15]. The absence of sideband peaks in Fig. 1c suggests that boron dominantly forms $sp^2$-bonds with nitrogen[16–18]. The h-BN layer is transferred onto transmission electron microscopy (TEM) grids for high-resolution imaging. The plan-view TEM image in Fig. 1e shows that the h-BN layer is mostly nanocrystalline or amorphous due to the low growth temperature. We next show graphene growth directly on GaAs(001) wafers in a III-V MOCVD system at a low temperature. Because the surface dissociation of GaAs occurs at an even lower temperature than GaN, a thin AlGaAs layer with a nominal Al composition of 50% ($Al_{0.5}Ga_{0.5}As$) is first grown on GaAs as a buffer that is thermally more robust and lattice-matched to GaAs. Next, using toluene as a carbon precursor with a low cracking temperature[19], graphene is formed on AlGaAs/GaAs substrate at around 700 °C. During the growth of graphene, toluene is exclusively flown to the reactor with a nitrogen carrier gas, while all group III and V precursors are turned off. A similar set of characterization is conducted on graphene as h-BN, and the AFM measurements reveal atomically smooth graphene/AlGaAs/GaAs surface with the r.m.s. roughness of 0.21 nm, as shown in Fig. 1f. The XPS spectra at C 1s peak reveal the peak binding energy of ~285 eV without notable sideband peaks, revealing that carbon lattices are mostly $sp^2$-bonded. The Raman spectra of graphene, which are measured by transferring the graphene onto Si substrates with a 300 nm-thick $SiO_2$ layer, show signature graphene peaks (G and 2D peaks) but with significant peak broadening. The slightly broad C 1s peak in XPS spectra and the broad 2D peak in Raman spectra agree perfectly with other reports on monolayer amorphous $sp^2$ carbon[20,21], which is also grown at a low temperature but on different substrates. This strongly suggests that the graphene layer formed on AlGaAs/GaAs is also an amorphous $sp^2$ phase. The plan-view TEM image of graphene in Fig. 1i, which is measured by transferring graphene onto TEM grids, also confirms the amorphous phase of graphene. In summary, h-BN and graphene are formed on III-N and III-V substrates, respectively, by a low-temperature growth.

With the capability to form 2D layers on compound semiconductor wafers in epitaxy systems, a completely new 'one-shot process' of remote epitaxy becomes possible, wherein 2D layers and epilayers are grown in a single growth run, without the need for 2D layer transfer, sample loading/unloading, or even temperature ramping up/down. Remote epitaxy of GaN can now be conducted by first forming the h-BN layer on a GaN wafer at 680 °C and then directly ramping the temperature up to 700 °C for GaN remote epitaxy. Similarly, AlGaAs buffer, graphene, and remote epitaxial GaAs layer can all be grown by a single MOCVD run. If necessary, it is possible to grow h-BN and GaN in a separate growth run, as well as to grow AlGaAs, graphene, and GaAs separately, allowing step-by-step characterizations. After the growth of GaN and GaAs thin films by such remote epitaxy processes respectively on 2-inch GaN(0001) and 2-inch GaAs(001) wafers, the material properties of remote epitaxial GaN and GaAs are characterized. The cross-sectional TEM image in Fig. 2a clearly shows the presence of h-BN interlayer between the GaN epilayer and the substrate, with the alignments of atomic structures of GaN through h-BN. The plan-view scanning-electron microscope (SEM) image in Fig. 2b shows the smooth surface of a 500 nm-thick GaN film. The electron backscatter diffraction (EBSD) map of the top surface reveals that the GaN layer is single-crystalline. The same set of measurements conducted on a 3 µm-thick GaAs film also reveals that a perfectly single-crystalline GaAs(001) film is formed on graphene-coated AlGaAs/GaAs substrate, as shown in Fig. 2e-g. The cross-sectional scanning TEM (STEM) image (Fig. 2e) shows a fully-merged graphene layer that is necessary for damage-free exfoliation of epilayers, where the full coverage of 2D layers is further demonstrated in the later section. Next, the grown films are exfoliated by 2DLT processes. A nickel stressor layer is deposited on the epilayers with a thin titanium adhesion layer, and then a thermal release tape (TRT) is attached to the stressor/epilayer/2D/substrate stack. Finally, the TRT/stressor/epilayer is mechanically released from the wafer. The photographs in Fig. 2d,h show exfoliated 2-inch GaN and GaAs membranes, confirming that the directly grown 2D interlayers are indeed an effective platform for remote epitaxy and 2DLT for single-crystal membrane production. The non-uniform color of the GaN membrane in Fig. 2d is due to the fluctuation of the thickness of GaN layer in our MBE growth environment, and is not related to remote epitaxy or h-BN interlayer.

One of the most substantial advantages of the utilization of directly-grown 2D layers on compound semiconductor substrates is their tearing-free and residue-free formation on entire wafers. When 2D materials such as graphene are transferred from foreign templates, it is *i*) difficult to cover the entire area of round-shape wafers, *ii*) transfer process-related residues and wrinkles deteriorate the quality of remote epitaxy[22,23], and *iii*) most importantly, tearing of 2D materials inevitably occurs in microscopic scales, from which direct epitaxy takes place that spalls the substrate during the exfoliation process by 2DLT[9]. All these challenges can be overcome by directly-grown 2D layers, as evidenced by the successful release of the full 2-inch-scale membranes shown in Fig. 2d,h. Furthermore, as the wafers remain pristine without any damage, it is possible to reuse the wafer after peeling without the need for thinning/planarizing the wafer by CMP, which makes a clear contrast to the conventional chemical, optical, and mechanical lift-off processes. In our approach, as schematically illustrated in Fig. 3a, GaAs wafer is reused after peeling of the remote epitaxial GaAs membrane, first by removing the remaining graphene by $O_2$ plasma, followed by a standard wafer cleaning and regrowth of graphene. Because AlGaAs buffer is already formed on the GaAs wafer, graphene is directly grown on the buffer. As a proof of concept, we utilize the same GaAs wafer three times to produce three 2-inch GaAs membranes. The photographs in Fig. 3b show three GaAs membranes and the donor wafer after each peeling. Each 2DLT process is conducted by the same process of Ti/Ni deposition, TRT attachment, and peeling from the edge. The wafer surface characterized by SEM and AFM after each peeling (Fig. 3c,d) shows no degradation of substrates throughout the recycling process, in both macroscopic and microscopic scales. Spalling of the substrates is not observed in the low-magnification SEM images, and the surface roughnesses measured by AFM are all around 1 nm after three times of remote epitaxy and 2DLT processes (r.m.s. roughness of 0.74 nm, 0.81 nm, and 1.02 nm after each exfoliation, in Fig. 3d). The slight damage near the upper edge of the wafer, which can be observed in the photographs in Fig. 3b, is owing to an edge effect in the growth of graphene in our MOCVD reactor. EBSD maps of exfoliated membranes in Fig. 3e show that the interface sides of the epilayers are all single-crystalline, not just the top surfaces of the epilayer (Fig. 2g), corroborating the "remote" nucleation through graphene. The X-ray diffraction (XRD) spectra of 2θ-ω scan for three exfoliated membranes (Fig. 3f) also confirm that the entire membranes are purely (001)-oriented, showing the peaks from only GaAs(001), Ti, and Ni. The reusability of GaN wafers is also demonstrated by the same process, *i.e.* 2DLT of GaN/h-BN layer followed by removing the remaining h-BN layer on the surface by $O_2$ plasma (data not shown). Therefore, directly-grown 2D materials on compound semiconductor substrates work as an effective template for wafer-scale remote epitaxy and wafer recycling, as the substrate damage is prevented during the exfoliation.

The capability to directly form 2D materials on the substrates of interest enables the growth of consecutive layers of 2D materials and 3D bulk layers by a single growth run, without the need for 2D layer transfer steps. The maximum number of such 2D/3D stacks that can be simultaneously grown is not limited in principle, and only constrained by the epitaxial systems and practical growth durations. By growing multiple remote epitaxial layers with 2D interlayers, it is possible to harvest many freestanding single-crystalline membranes by repeated mechanical exfoliation from the topmost layer, which is an extremely high-throughput process in terms of both the epitaxy and membrane harvest when compared with conventional layer lift-off techniques such as epitaxial lift-off and optical lift-off. Although an advanced epitaxial lift-off technique has been proposed by employing multiple sacrificial layers and transfer printing[24], isolation of each layer requires complex processes involving hole patterning/dry-etching, partial wet-etching, PR anchoring, and final release, which are time-consuming and less practical. Furthermore, since the wafer can be reused in our approach without the need for CMP as demonstrated above, it has a great potential for the reduction of manufacturing costs by reusing costly wafers. For a proof-of-concept demonstration, we first perform the growth of three stacks of GaN/h-BN on a 2-inch GaN(0001) wafer, by a single growth run in an MBE system. The XRD $\phi$-scan of the as-grown structure in Fig. 4a shows six-fold symmetry, which confirms that all three GaN epilayers are crystallographically aligned with the GaN substrate of a wurtzite crystal phase. The layer-by-layer peeling is performed in the same process as the single-layer case in Fig. 3, by depositing a Ti/Ni stressor layer and then peeling with a handling tape. It is worth mentioning that only the topmost layer is exfoliated per each 2DLT process, which is attributed to the full coverage of 2D interlayers on each GaN layer that facilitates seamless crack propagation along the 2D interface. Therefore, three membranes are individually harvested by three consecutive 2DLT processes. After each peeling, h-BN exposed on the top is removed by $O_2$ plasma before another deposition of Ti/Ni for the next 2DLT process. As shown in the cross-sectional and plan-view SEM images in Fig. 4b and 4c that are measured before and after each 2DLT process, the peeling occurs in a layer-by-layer manner with a smooth morphology of the surface after each exfoliation, confirming that each GaN epilayer is undamaged during the exfoliation of upper layers. The EBSD maps in Fig. 4c, which are measured from the interface side of exfoliated GaN membranes, also confirm the remote epitaxial growth mode of GaN epilayers through three h-BN interlayers. Next, we perform similar studies by growing three stacks of (Al)GaAs/graphene on a GaAs(001) substrate using a III-V MOCVD system. To prevent surface degradation, AlGaAs layer is first grown as a thermal buffer before graphene formation. As shown in Fig. 4d, the XRD $\phi$-scan of three GaAs stacks confirms four-fold symmetry of zinc-blende GaAs layers. All three layers are individually exfoliated by three consecutive 2DLT processes (Fig. 4e), and like the case of GaN/h-BN multilayers, graphene is removed by $O_2$ plasma after each peeling. The SEM images of substrates in Fig. 4f after each peeling show that each layer is exfoliated without any damage on the substrate, and the EBSD maps of exfoliated GaAs membranes prove the single-crystallinity of each epilayer. Therefore, these results exemplify the effectiveness of our approach in producing multiple freestanding membranes by a single growth run on a single wafer.

In conclusion, we have demonstrated a method for producing single-crystalline semiconductor membranes with an extremely high throughput potentially at a low cost. The key enabling technique is the formation of 2D materials directly on wafers in epitaxy systems, which allows for simultaneous growth of remote epitaxial films with 2D materials, and even multiple stacks of them. By combining this technique with remote epitaxy and 2DLT, single or multiple stacks of 3D/2D layers are grown and exfoliated by mechanical peeling, where the substrates can be reused after the peeling for further production of additional membranes. We envision that this technique can be applied not only to III-V/III-N compound semiconductors, but also to many other materials by developing 2D layer growth on them. Also, the available choice of materials can be further expanded if 2D layers can be grown at even lower temperatures. Indeed, there are active studies on the low-temperature growth of sp$^2$-bonded 2D layers, such as the growth of amorphous carbon and amorphous boron nitride at as low as 200 °C[20] and 400 °C[25], respectively. Therefore, this membrane production method could become a universal method to produce single-crystalline membranes with great flexibility in the choice of materials and the substrate, benefiting diverse application fields of heterointegrated electronics and functional device platforms.

**Competing interests**

The authors declare no competing interests.

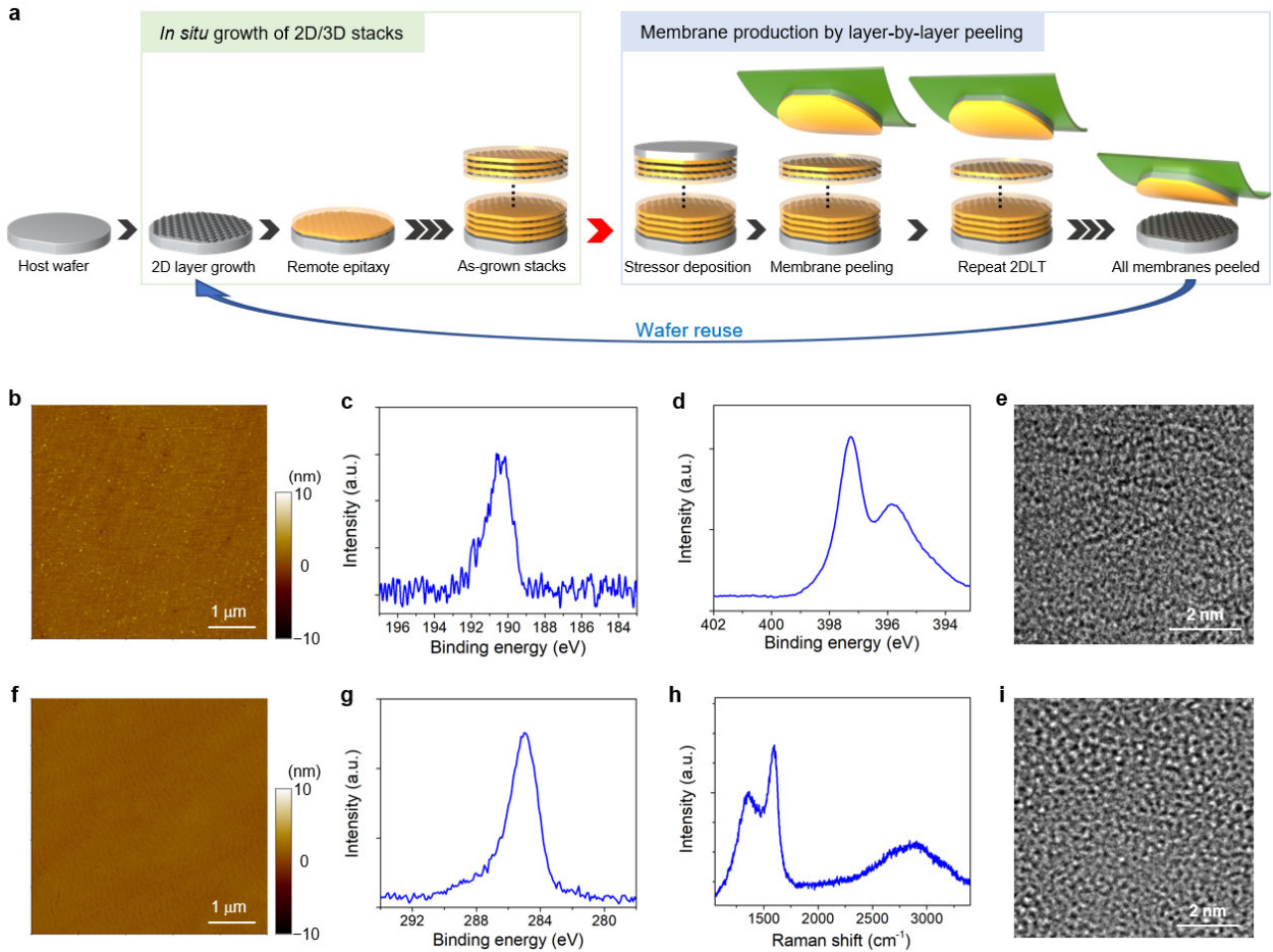

**Fig. 1 | Direct growth of 2D materials for advanced remote epitaxy and layer transfer. a**, Schematic of single-crystalline membrane production process by *in situ* growth of 2D/3D stacks and layer-by-layer peeling. **b-d**, AFM image of h-BN formed on GaN substrate (**b**) and XPS spectra measured around B 1s peak (**c**) and N 1s peak (**d**). **e**, Plan-view TEM image measured by transferring h-BN onto a TEM grid. **f,g**, AFM image of graphene formed on AlGaAs/GaAs substrate (**f**) and XPS spectra measured around C 1s peak (**g**). **h**, Raman spectra of graphene measured after transferring graphene onto SiO$_2$/Si substrate. **i**, Plan-view TEM image of graphene measured by transferring graphene onto a TEM grid.

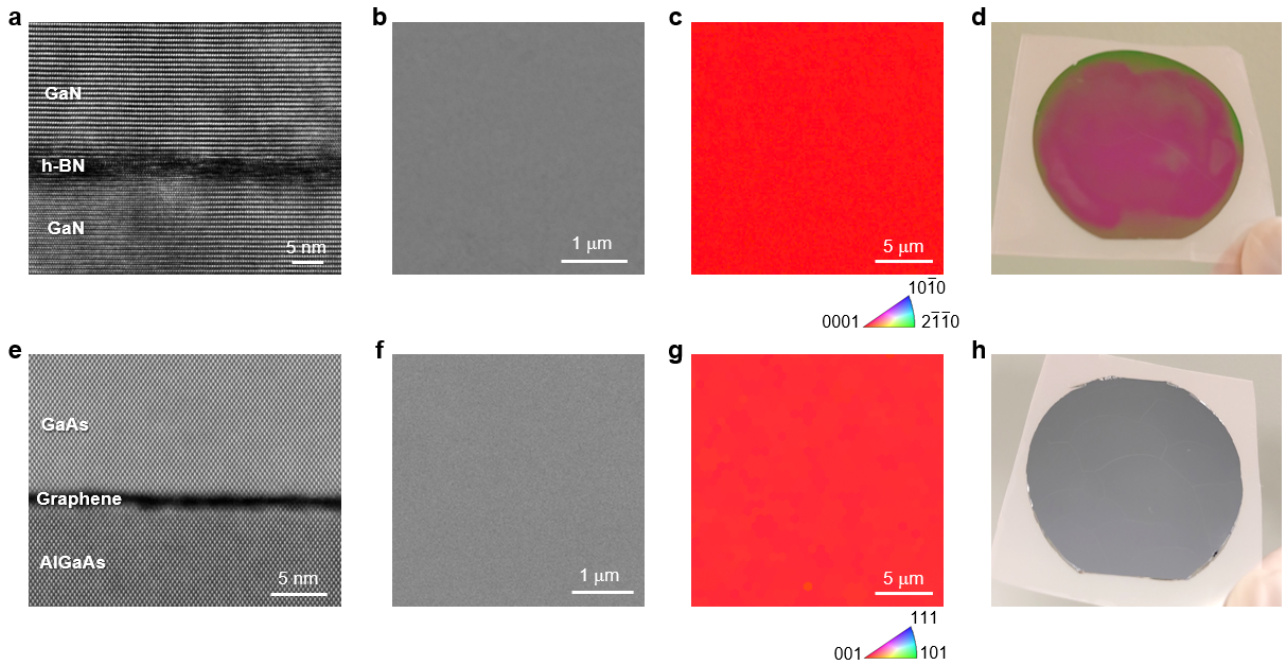

**Fig. 2 | Remote epitaxy on directly grown 2D materials. a-d**, Remote epitaxy of GaN on h-BN-coated 2-inch GaN wafer. Cross-sectional TEM image (**a**), plan-view SEM image (**b**), EBSD map (**c**) of remote epitaxial GaN showing single-crystallinity, and photograph of exfoliated 2-inch GaN membrane (**d**). **e-h**, Similar set of data for GaAs grown on graphene-coated 2-inch AlGaAs/GaAs wafer. Cross-sectional STEM image (**e**), plan-view SEM image (**f**), EBSD map (**g**) of as-grown sample, and photograph of exfoliated 2-inch GaAs membrane (**h**).

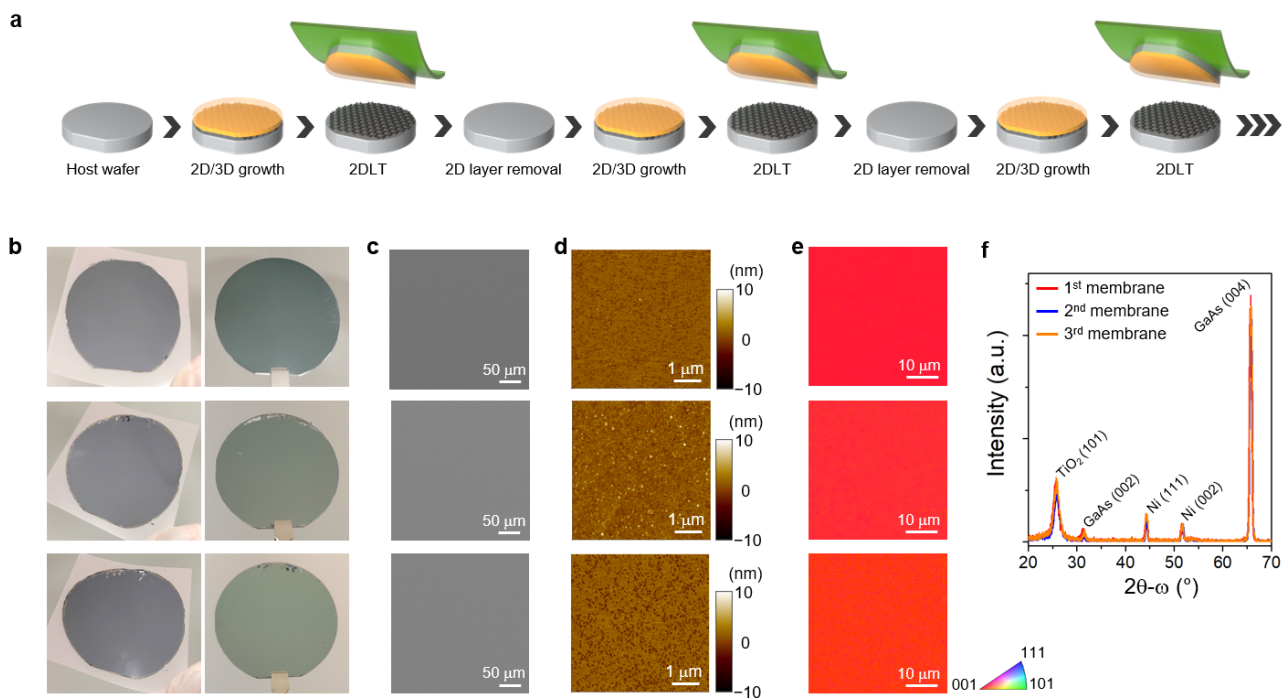

**Fig. 3 | Multiple membrane production by reusing a wafer. a**, Schematic of successive membrane production by 2DLT and wafer recycling. **b-e**, Characterizations of exfoliated GaAs membranes and the substrate after each growth and 2DLT. Photographs of exfoliated membranes (**b**, left) and the remaining substrate (**b**, right), SEM images (**c**) and AFM images (**d**) of the substrate, and EBSD maps of exfoliated GaAs membrane (**e**). The data in the upper, middle, and lower rows correspond to the first, second, and third growth and exfoliation from the same wafer, respectively. **f**, XRD spectra of exfoliated membranes.

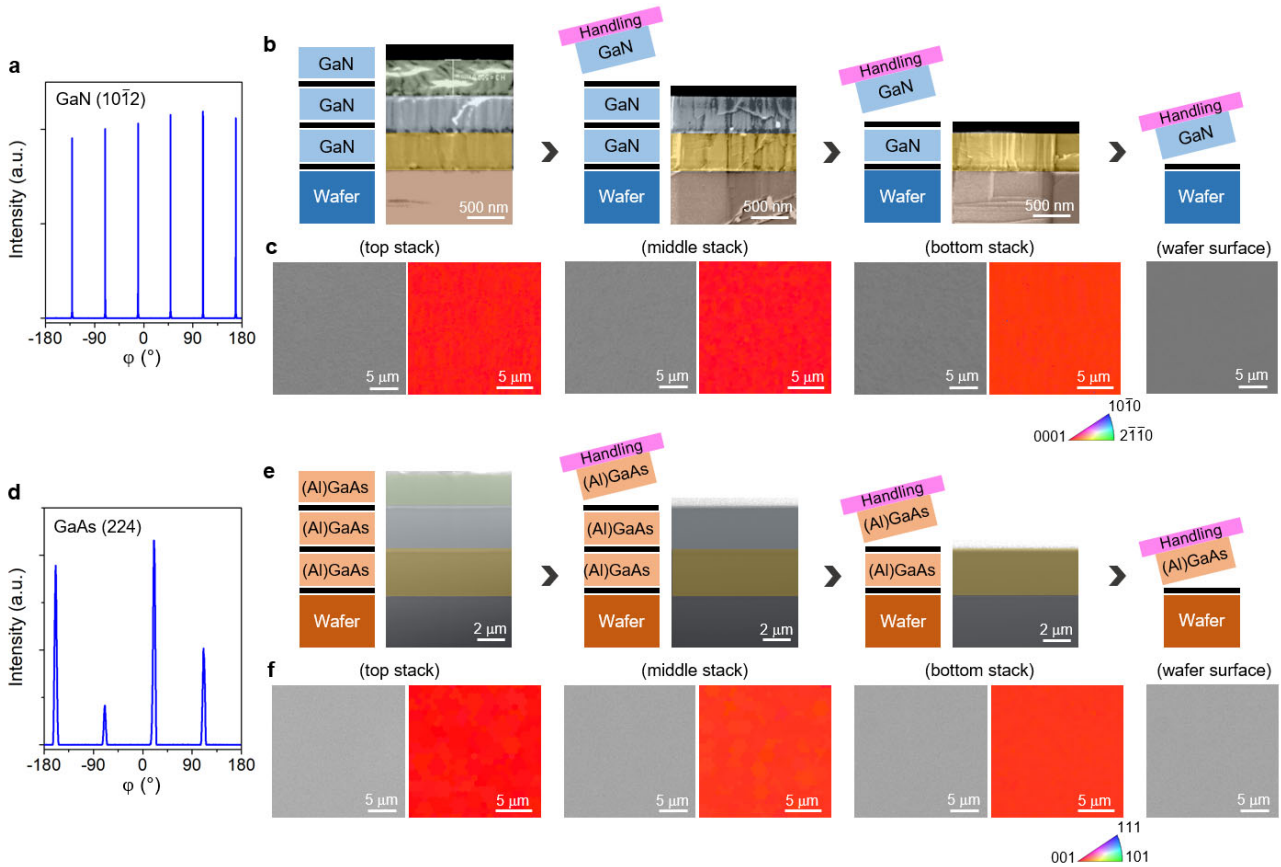

**Fig. 4 | Multi-stack growth and layer-by-layer exfoliation. a**, XRD spectra of three stacks of GaN/h-BN grown on a GaN substrate. **b**, Schematics and false-color cross-sectional SEM images of as-grown sample (left) and after each peeling (center, right). **c**, Plan-view SEM images of top surfaces (left) and EBSD maps of membranes measured after exfoliation (right). **d-f**, Same set of data for three stacks of AlGaAs/GaAs/graphene grown on an AlGaAs/GaAs substrate.

15